\documentclass[preprint,showpacs,preprintnumbers,amsmath,amssymb]{revtex4-1}
\usepackage{latexsym}
\usepackage{epsfig}
\usepackage{float}
\usepackage{lipsum}
\usepackage{graphicx,times}
\usepackage{amssymb,amsmath}
\usepackage{color}
\usepackage{dcolumn}
\usepackage{epstopdf}
\usepackage{pgfplots}
\usepackage[colorlinks=true, urlcolor=blue, linkcolor=blue, citecolor=blue]{hyperref}
\newcommand{\be}{\begin{equation}}
\newcommand{\ee}{\end{equation}}

\begin{document}
\title{Three- and four-body kaonic nuclear states: Binding energies and widths}
\author{S. Marri}
\affiliation{Department of  Physics, Isfahan University of Technology, Isfahan 84156-83111, Iran}
\author{J. Esmaili}
\affiliation{Department of Physics, Faculty of Basic Sciences, Shahrekord University, Shahrekord, 115, Iran}
\date{\today}
\begin{abstract}
Using separable potentials for $\bar{K}N-\pi\Sigma$ interaction, we 
investigated four-body kaonic nuclear systems such as $K^{-}ppn$ 
and $K^{-}K^{-}pp$, with the Faddeev AGS method in the momentum 
representation. The Faddeev calculations are based on the quasi-particle 
method and the method of the energy dependent pole expansion was used 
to obtain the separable representation for the integral kernels in the 
three- and four-body equations. Different types of $\bar{K}N-\pi\Sigma$ 
potentials based on phenomenological and chiral SU(3) approach are used 
and it was shown that the kaonic nuclear systems under consideration are 
tightly bound. 
\end{abstract}
\maketitle
\section{Introduction}
\label{intro}
The $\bar{K}N$ interaction, which is affected by $\Lambda$(1405) resonance, 
plays an important role in the exotic systems, including the antikaon 
particle~\cite{dal1,dal2,akaishi,esm1,esm2}. Thus, to study the kaonic systems, 
it is necessary to know the $\bar{K}N$ interaction. The first prediction of a 
quasi-bound state in kaonic nuclear systems was made in~\cite{akaishi,yamazaki1,dot1,dot2}, 
showing that these systems could be strongly bound. For the past two decades, many 
theoretical calculations were performed, focusing on the three- and four-body 
kaonic systems~\cite{shev1,shev2,ikeda1,ikeda2,dote1,dote2,ikeda3,kh1,kh2,kh3,maeda,gal1}.

Alongside theoretical studies, many experimental searches have been also 
carried out to investigate the possible existence of the quasi-bound state 
in the kaonic systems (especially $K^{-}pp$ system). The investigations for 
the $K^{-}pp$ quasi-bound state have been explored by FINUDA experiment at 
the DAPhNE collider~\cite{agnel} and also by OBELIX at CERN~\cite{bend} and 
DISTO at SATURNE~\cite{yamaz}. Further experimental results were obtained by 
E15 and E27 groups at J-PARC~\cite{aji,ichi}. However, the possible existence 
of the quasi-bound state in the $K^{-}pp$ systems is still highly uncertain 
and there are some doubts in the extracted experimental results. The new 
planned experiments by HADES~\cite{fabbi} and LEPS~\cite{toki} Collaborations, 
and also by J-PARC~\cite{aji,ichi} experiments may unravel this problem.

The purpose of the present paper is to explore the binding energy and width 
of four-body kaonic nuclear systems including one or two antikaon particle. 
The problem can be solved using methods developed within four-body theories. 
To reduce the four-body Faddeev equations to a set of single-variable integral 
equations, one can employ different methods~\cite{naro,sofia}. One can do the 
reduction procedure numerically by making use of the so-called HSE method 
proposed by Narodetsky~\cite{naro} and also by using the energy-dependent pole 
expansion method which developed by Sofianos {\it et al}~\cite{sofia}. In 
Refs~\cite{kh1} and~\cite{kh2}, the HSE method was employed to solve the Faddeev 
equations of $K^{-}ppn$ and $K^{-}K^{-}pp$ systems, respectively. One can also 
perform the four-body calculation using the energy-dependent pole expansion 
method~\cite{sofia} or the so-called EDPE method. In EDPE method the form 
factors are energy dependent. In the present study, $K^{-}ppn$ and $K^{-}K^{-}pp$ 
quasi-bound state positions were calculated. Using the EDPE method, we found 
the separable expressions for the [3+1] and [2+2] subsystems. At the same time, 
the obtained results for EDPE method can be compared with those by Hilbert-Schmidt 
pole expansion methods and also study the behavior of the binding energy and 
width of kaonic systems under these situations. The dependence of the pole energy 
on different models of $\bar{K}N-\pi\Sigma$ interaction will be studied. 

There is an opinion that the $K^{-}pp$ system has a two-pole structure similar 
to the $\bar{K}N$ system~\cite{dot3}. To study this issue, the Faddeev amplitudes 
for $\bar{K}N$ and $\bar{K}NN$ systems were calculated in the complex energy plane. 
With this method, we investigated how the pole energy manifest itself in two- and 
three-body scattering amplitudes. We examined whether the first and the second pole 
of these systems can be seen in the corresponding scattering amplitudes. Different 
models of interactions, which are derived chirally and phenomenologically, will be 
included in our calculations for $\bar{K}N-\pi\Sigma$ system~\cite{ikeda4,shev3}.

The paper is organized as follows: in Sect.~\ref{formula}, we will explain the 
formalism used for the four-body $\bar{K}NNN$ and $\bar{K}\bar{K}NN$ systems and 
give a brief description of the quasi-particle method and separable representation 
of the Faddeev amplitudes by EDPE method. The two-body inputs of the calculations 
and the computed binding energies and widths are presented in Sect.~\ref{result} 
and in Section~\ref{conc}, we give conclusions.\\
\section{Three- and four-body calculations}
\label{formula}
In the present work, the possible existence of a quasi-bound state in the 
$K^{-}ppn$ and $K^{-}K^{-}pp$ four-body systems was studied. We used the quasi-particle 
method to solve the four-body Faddeev equations. The key point of the quasi-particle 
method is the separable representation of the off-shell scattering amplitudes in 
two- and three-body subsystems~\cite{naro,grass,fonce}. Using properly symmetrized and 
antisymmetrized states with respect to identical kaons and nucleons, we will have the 
following subsystems of the $\bar{K}NNN$ four-body systems, without defining the 
interacting pairs. 
\begin{equation}
\begin{split}
& \alpha=1:\bar{K}+(NNN), \ \ \alpha=2:N+(\bar{K}NN),\\
& \alpha=3:(\bar{K}N)+(NN),
\end{split}
\label{eq1}
\end{equation}

The quantum numbers of the $\bar{K}NNN$ are $I=0$ and $s=\frac{1}{2}$, in actual 
calculations, when we include isospin and spin indexis the number of configurations 
is equal to twelve, corresponding to different possible two-quasi-particle partitions.
\begin{equation}
\begin{split}
& \bar{K}(N[NN]_{s=0,1}), \,\, (\bar{K}[NN]_{s=0,1})N, \\
& ([\bar{K}N]_{I=0,1}N)_{s=0,1}N,  \\
& [\bar{K}N]_{I=0,1}+NN, \,\, [NN]_{s=0,1}+\bar{K}N.
\end{split}
\label{eq55}
\end{equation}

In the case of $\bar{K}\bar{K}NN$ system, we have one pair of identical kaon and 
one pair identical nucleon. Therefore, we will have four different subsystems, 
which are given by
\begin{equation}
\begin{split}
& \alpha=1:\bar{K}+(\bar{K}NN), \ \ \alpha=2:N+(\bar{K}\bar{K}N),\\
& \alpha=3:(\bar{K}N)+(\bar{K}N), \ \ \alpha=4:(\bar{K}\bar{K})+(NN).
\end{split}
\label{eq2}
\end{equation}
the quantum numbers of the $\bar{K}\bar{K}NN$ are $I=0$ and $s=0$. Therefore, the 
number of configurations will be ten when we add the isospin and spin indexis  
\begin{equation}
\begin{split}
& \bar{K}(\bar{K}[NN]_{s=0}), \,\, \bar{K}([\bar{K}N]_{I=0,1}N), \\
& ([\bar{K}\bar{K}]_{I=1}N)N, \,\, (\bar{K}[\bar{K}N]_{I=0,1})N, \\
& [\bar{K}\bar{K}]_{I=1}+NN, \,\, \bar{K}\bar{K}+[NN]_{s=0}, \\
& [\bar{K}N]_{I=0,1}+\bar{K}N.
\end{split}
\label{eq66}
\end{equation}

\begin{figure*}[htb]
\vspace{ 0cm}
\centering
\includegraphics[scale=0.55]{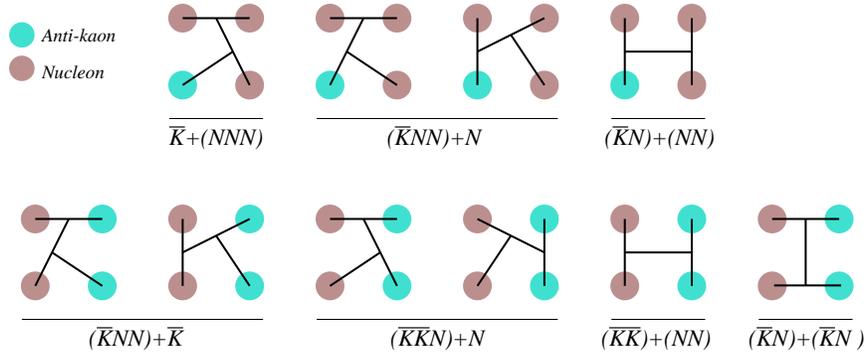}
\vspace{-0cm}
\caption{ (Color online) Diagrammatic representation of different 
partitions of the $\bar{K}NNN$ (up) and $\bar{K}\bar{K}NN$ (down) 
systems without including the particle, spin and isospin labels. 
The anti-kaons are defined by turquoise circles and the nucleons by 
brown circles.} 
\label{fig1}
\end{figure*}

The whole dynamics of $\bar{K}NNN$ system is described in terms of the transition 
amplitudes $\mathcal{A}_{\alpha\beta}$ which connect the quasi-two-body channels 
characterized by Eqs. (\ref{eq1}) and (\ref{eq2}). In Fig.~\ref{fig1}, the four 
different rearrangement channels of the $\bar{K}NNN$ and six rearrangement channels 
of the $\bar{K}\bar{K}NN$ four-body system including the K- and H-type diagrams are 
represented. In Fig.~\ref{fig1}, the partitions defined in~\ref{eq1} and~\ref{eq2} 
are depicted, including the two-quasi-particles in the subsystems. Antisymmetrization 
of nucleons and symmetrization of the kaons to be made within each channel. The Faddeev 
equations for kaonic systems under consideration can be expressed by~\cite{fix,naka}
\begin{equation}
\begin{split}
& \mathcal{A}^{I_{i}I_{j},ss'}_{\alpha(i)\beta(j),nn'}(p,p',E)=
\mathcal{R}^{I_{i}I_{j},ss'}_{\alpha(i)\beta(j),nn'}(p,p',E) \\
& \hspace{1.cm}+\sum_{\gamma;lm}\sum_{I_{k},s''}\int{d}\vec{p}''\mathcal{R}^{I_{i}I_{k},ss''}_{\alpha(i)
\gamma(k),nl}(p,p'',E)\,\theta^{\gamma,s''}_{lm}\, \\
& \hspace{1.cm}\times\mathcal{A}^{I_{k}I_{j},s''s'}_{\gamma(k)\beta(j),mn'}(p'',p',E).
\end{split}
\label{eq3}
\end{equation}

Here, the operators $\mathcal{A}^{I_{i}I_{j},ss'}_{\alpha(i)\beta(j),nn'}$ are 
the four-body transition amplitudes, which describe the dynamics of the four-body 
$\bar{K}NNN$ and $\bar{K}\bar{K}NN$ systems and the $\theta^{\gamma,s}_{lm}$-functions 
are the effective propagators. The total energy of the four-body system and the 
momentum of the spectator particle is defined by $E$ and $p$, respectively. To define 
the spectator particle or interacting particles in each two- and three-body subsystem, 
we used the $i, j$ and $k$ indices and the isospin of the interacting particles are 
defined by $I_ {i}$. The indices $n,l,m$ are used for defining which term of the 
separable expansion of the subamplitudes is used. The operators 
$\mathcal{R}^{I_{i}I_{j},ss'}_{\alpha(i)\beta(j),nn'}$ are driving terms, which describe 
the effective particle-exchange potential realized by the exchanged particle between the 
quasi-particles in channels $\alpha$ and $\beta$, which can be written as

\begin{equation}
\begin{split}
& \mathcal{R}^{I_{i}I_{j},ss'}_{\alpha(i)\beta(j),nn'}(p,p',E)=\frac{\Omega^{I_{i}I_{j}}_{ss'}}{2}
\int^{+1}_{-1}d(\hat{p}\cdot\hat{p}')\\
& \hspace{0.8cm}\times{u}^{\alpha,s}_{n,iI_{i}}(\vec{q},\epsilon_{\alpha}-
\frac{p^{2}}{2\mathcal{M}^{\alpha}_{i}})\tau(z)
u^{\beta,s'}_{n',jI_{j}}(\vec{q}',\epsilon_{\beta}-\frac{p'^{2}}{2\mathcal{M}^{\beta}_{j}}),
\end{split}
\label{eq4}
\end{equation}
where the symbols $\Omega^{I_{i}I_{j}}_{ss'}$ are the Clebsch-Gordan coefficients, 
the functions ${u}^{\alpha,s}_{n,iI_{i}}$ are the form factors that generated by 
the separable representation of the sub-amplitudes appearing in the channels 
(\ref{eq1} and \ref{eq2}) and $z$ is given as 
$z=E-\frac{p^{2}}{2M^{\beta}_{j}}-\frac{p'^{2}}{2M^{\alpha}_{i}}-\frac{\vec{p}\cdot\vec{p}'}{m}$.
The subsystem energies are defined by $\epsilon_{\alpha}$. Finally, the momenta 
$\vec{q}(\vec{p},\vec{p}')$ and $\vec{q}'(\vec{p},\vec{p}')$ are given in terms 
of $\vec{p}$ and $\vec{p'}$. We use the relations
\begin{equation}
\vec{q}=\vec{p}'+\frac{M^{\alpha}_{i}}{m}\vec{p}, \hspace{1cm} 
\vec{q}'=\vec{p}+\frac{M^{\beta}_{j}}{m}\vec{p}',
\label{eq5}
\end{equation}
where $m$ is exchanged particle or quasi-particle mass and the reduced masses 
$\mathcal{M}_{\alpha}$and $M_{\alpha}$ in the channel $\alpha$ of the [3+1] 
subsystem are defined by
\begin{equation}
\begin{split}
& \mathcal{M}^{\alpha}_{i}= m^{\alpha}_{i}(m^{\alpha}_{j}+m^{\alpha}_{k}+m^{\alpha}_{l})
/(m^{\alpha}_{i}+m^{\alpha}_{j}+m^{\alpha}_{k}+m^{\alpha}_{l}), \\
& M^{\alpha}_{j} = m^{\alpha}_{j}(m^{\alpha}_{k}+m^{\alpha}_{l})
/(m^{\alpha}_{j}+m^{\alpha}_{k}+m^{\alpha}_{l}),
\end{split}
\label{eq6}
\end{equation}

Before solving the four-body equations, one should solve the bound 
state problem for the two- and three-body subsystems that are specified 
in the partitions (\ref{eq1} and \ref{eq2}). The three-body Faddeev 
equations~\cite{shev2} in the AGS take the form 
\begin{equation}
\mathcal{K}_{ij,I_{i} I_{j}}^{\alpha,s}=\mathcal{M}_{ij,I_{i} I_{j}}^{\alpha,s}+\sum_{k,I_{k}}
\mathcal{M}_{ik,I_i I_k}^{\alpha,s}\tau_{k,I_k}^{\alpha,s}\mathcal{K}_{kj,I_k I_j}^{\alpha,s}.
\label{eq8}
\end{equation}

The inputs for the AGS system of equations (\ref{eq8}) are the two-body 
operators $\tau_{k,I_k}^{\alpha,s}$, evaluated in the presence of a spectator 
particle. The operators $\mathcal{K}_{ij,I_{i} I_{j}}^{\alpha,s}$ are the 
usual transition amplitudes between Faddeev channels~\cite{shev2} and the 
operators $\mathcal{M}^{\alpha,s}_{ij,I_{i}I_{j}}$ are the corresponding 
Born terms. Faddeev partition indices $i,j,k=1,2,3$ denote simultaneously 
an interacting pair and a spectator particle. 

To take the coupling between $\bar{K}N$ and $\pi\Sigma$ channels directly 
into account, the formalism of Faddeev equations should be extended to 
include the particle channels~\cite{shev2,ikeda1}. Thus, all three-body 
operators should have particle indices for each state in addition to the 
Faddeev indices. In the present calculations, the $\pi\Sigma{N}$ channel 
of the  $\bar{K}NN$ system and $\pi\bar{K}\Sigma$ channel of the $\bar{K}\bar{K}N$ 
system have not been included directly and one-channel Faddeev AGS equations 
are solved for the $\bar{K}NN$ and the $\bar{K}\bar{K}N$ systems. We approximated 
the full coupled-channel interaction by constructing the so-called exact optical 
$\bar{K}N-\pi\Sigma$ potential~\cite{shev4}. The exact optical potential provides 
exactly the same elastic $\bar{K}N$ scattering amplitude as the coupled-channel 
model of interaction. Thus, our coupled-channels four-body calculations with 
coupled-channel $\bar{K}N-\pi\Sigma$ interaction is equivalent to the one-channel 
four-body calculation using the so-called exact optical $\bar{K}N(-\pi\Sigma)$ 
potential. The decaying to the $\pi\Sigma{N}$ and $\pi\bar{K}\Sigma$ channels is 
taken into account through the imaginary part of the optical $\bar{K}N(-\pi\Sigma)$ 
potential. Since, we do not include the $\pi\Sigma{N}$ and $\pi\bar{K}\Sigma$ 
channels directly into our calculations, in Eq. (\ref{eq8}) we neglected the 
particle indices of the operators.

We have to introduce a separable representation for the three-body 
amplitudes and driving terms, which will be necessary to find the 
solution of three-body subsystems. In the present work, for this purpose 
we apply the EDPE expansion method~\cite{naro,fix}. The separable form 
of the Faddeev transition amplitudes is given by
\begin{equation}
\mathcal{K}_{ij,I_i I_j}^{\alpha,s}(q,q',\epsilon)=\sum_{n,m}^{N_{r}} 
u^{\alpha,s}_{n,iI_{i}}(q,\epsilon) 
\theta^{\alpha,s}_{nm}(\epsilon) u^{\alpha,s}_{m,jI_{j}}(q',\epsilon).
\label{eq9}
\end{equation} 

The starting point of the energy dependent pole expansion method (EDPE) 
is the eigenvalue equations for the vertex functions 
$u^{\alpha,s}_{n,iI_{i}}(q,B_{\alpha})$ 
\begin{eqnarray}
\begin{split}
& u^{\alpha,s}_{n,iI_{i}}(q,B_\alpha)=\frac{1}{\lambda^{\alpha}_n}
\sum\limits_{jI_{j}}\int \mathcal{M}^{\alpha,s}_{ij,I_{i}I_{j}}(q,q';B_\alpha)\, \\
& \hspace{2cm}\times \tau^{\alpha,s}_{jI_{j}} \big(B_\alpha-\frac{{q'}^2}{2M^{\alpha}_{j}}\big) 
u^{\alpha,s}_{n,jI_{j}}(q',B_\alpha)d\vec{q}^{\prime}.
\end{split}
\label{eq10}
\end{eqnarray}

By solving equations (\ref{eq8}), we can define the binding energy and width of 
${K^{-}pp}$, $K^{-}K^{-}p$ and $^{3}{He}$ systems. Since, the $K^{-}d$ is not 
bound, in Eq.~(\ref{eq8}), we will put $B_{K^{-}d}=-\epsilon_{d}$ (the deuteron 
binding energy). Therefore, by taking $B_{K^{-}pp}=-\epsilon_{K^{-}pp}$ (the 
$K^{-}pp$ binding energy), $B_{K^{-}K^{-}p}=-\epsilon_{K^{-}K^{-}p}$ (the 
$K^{-}K^{-}p$ binding energy) and $B_{NNN}=-\epsilon_{^{3}{He}}$ (the triton 
binding energy), we can define the form factors $u^{\alpha,s}_{n,iI_{i}}(q,B_\alpha)$ 
for each state. The extrapolation of the vertices $u^{\alpha,s}_{n,iI_{i}}$ onto the 
whole energy axes is achieved according to the following expression
\begin{eqnarray}
\begin{split}
& u^{\alpha,s}_{n,iI_{i}}(q,\epsilon)=\frac{1}{\lambda^{\alpha}_n}
\sum\limits_{jI_{j}}\int\mathcal{M}^{\alpha,s}_{ij,I_{i}I_{j}}(q,q';\epsilon)\, \\
& \hspace{2cm}\times \tau^{\alpha,s}_{jI_{j}} \big(B_\alpha-\frac{{q'}^2}
{2M^{\alpha}_{j}}\big) 
u^{\alpha,s}_{n,jI_{j}}(q',B_\alpha)
d\vec{q}^{\prime}.
\end{split}
\label{eq11}
\end{eqnarray}

After finding the vertex functions $u^{\alpha,s}_{n,iI_{i}}(q,\epsilon)$, 
we can define the effective EDPE propagators $\theta_{\alpha}(\epsilon)$ 
in Eqs.~\ref{eq3} and~\ref{eq9} by
\begin{eqnarray}
\begin{split}
& \big(\theta^{(s)-1}_\alpha(\epsilon)\big)_{mn}=\sum_{jI_{j}}\int
\big[u^{\alpha,s}_{m,jI_{j}}(q,B_\alpha)\tau^{\alpha,s}_{jI_{j}}
\big(B_\alpha-\frac{q^2}{2M^\alpha_j}\big) \\
& -u^{\alpha,s}_{m,jI_{j}}(q,\epsilon)\tau^{\alpha,s}_{jI_{j}}
\big({\epsilon}-\frac{q^2}{2M^\alpha_j}\big)
\big] u^{\alpha,s}_{n,jI_{j}}(q,\epsilon)d\vec{q}.
\end{split}
\label{eq12}
\end{eqnarray}

Before we proceed to solve the four-body equations, we also need as 
input the equations describing two independent pairs of interacting 
particles such as $(\bar{K}N)(NN)$, $(\bar{K}\bar{K})(NN)$ and 
$(\bar{K}N)(\bar{K}N)$~\cite{kh1,kh2}. Thus, one should define the 
vertex functions and the EDPE propagators for each isospin state of 
these subsystems. We have taken 
$B_{(\bar{K}N)_{I=0}(NN)}=-\epsilon_{(\bar{K}N)_{I=0}}$ (the $\Lambda(1405)$ 
resonance mass and width), $B_{(\bar{K}N)_{I=1}(NN)}=-\epsilon_{d}$, 
$B_{(\bar{K}\bar{K})_{I=1}(NN)}=-\epsilon_{d}$, 
$B_{(\bar{K}N)_{I=0}(\bar{K}N)_{I=0}}=-\epsilon_{(\bar{K}N)_{I=0}}$ and 
finally $B_{(\bar{K}N)_{I=1}(\bar{K}N)_{I=1}}=0$.
\section{RESULTS AND DISCUSSION}
\label{result}
Before we proceed to represent the obtained results, we will have a survey 
on the two-body interactions. The two-body interactions are the central 
input to our few-body calculations. The orbital angular momentum of all 
interactions is taken to be zero. We used separable potentials in momentum 
representation in the form
\begin{equation}
V_{I}^{\alpha\beta}(k^{\alpha},k^{\beta};E)=g_{I}^{\alpha}(k^{\alpha})
\lambda_{I}^{\alpha\beta}g_{I}^{\beta}(k^{\beta}),
\label{eq13}
\end{equation}
where $g_{I}^{\alpha}(k^{\alpha})$ is the form factor of the interacting 
two-body system with relative momentum $k^{\alpha}$ and isospin $I$. Here, 
$\lambda_{I}^{\alpha\beta}$ is the strength parameter of the interaction. 
The interactions are further labeled with the $\alpha$ values to take the 
$\bar{K}N-\pi\Sigma$ coupling directly into account. Using separable potentials 
in the form~\ref{eq13} for two-body interaction, we can define the two-body 
t-matrices in the form
\begin{equation}
T_{I}^{\alpha\beta}(k^{\alpha},k^{\beta};E)=g_{I}^{\alpha}(k^{\alpha})
\tau_{I}^{\alpha\beta}(E)g_{I}^{\beta}(k^{\beta}),
\label{eq14}
\end{equation}
where the operator $\tau_{I}^{\alpha\beta}(E)$ is the usual two-body 
propagator. To describe the $\bar{K}N-\pi\Sigma$ interaction, which 
plays a crucial role in the present three- and four-body calculations, 
we considered three different phenomenological and chiral 
potentials~\cite{ikeda4,shev3}. The potentials have one- and two-pole 
structure of the $\Lambda(1405)$. The parameters of the $\bar{K}N-\pi\Sigma$ 
phenomenological potentials, are given in Ref.~\cite{shev3}. These 
potentials are adjusted to reproduce the SIDDHARTA experiment results~\cite{sidd}. 
Thus, depending on a pole structure of the $\Lambda(1405)$, we refer these 
potentials as \textquotedblleft{SIDD}-1\textquotedblright and 
\textquotedblleft{SIDD}-2\textquotedblright potential. The parameters 
of the $\bar{K}N-\pi\Sigma$ chiral potential, are given in Ref.~\cite{ikeda4} 
which is an energy-dependent potential. Another important interaction in our 
few-body calculatios is the nucleon-nucleon interaction. The potential that 
we considered here is the one-term PEST potential from Ref.~\cite{pest}, which 
is a separable approximation of the Paris model of $NN$ interaction. The 
parameters of the PEST potential are given in Ref.~\cite{pest}. 

The experimental information on $\bar{K}$-$\bar{K}$ interaction is poor. We 
used a separable potential for the $\bar{K}\bar{K}$ with $I=1$, in a Yamaguchi form
\begin{equation}
\begin{split}
& V^{I=1}_{\bar{K}\bar{K}}(k,k')=\lambda^{I=1}_{\bar{K}\bar{K}}g_{\bar{K}\bar{K}}(k)
g_{\bar{K}\bar{K}}(k'), \\
& \ \ \ \ \ \ \ g_{\bar{K}\bar{K}}(k)=\frac{1}{k^{2}+\Lambda^{2}_{\bar{K}\bar{K}}}.
\end{split}
\end{equation}

The range parameter value 3.9 $\mathrm{fm}^{-1}$ is adopted for $\bar{K}\bar{K}$ 
interaction to represent the exchange of heavy mesons and the strength parameter 
$\lambda^{I=1}_{\bar{K}\bar{K}}$ is adjusted to reproduce the $K^{+}K^{+}$ scattering 
length, for which we used as a guideline the result of lattice QCD calculation as 
$a_{K^{+}K^{+}}=−0.141$ fm~\cite{bean}.

It has been suggested in Ref.~\cite{dot3}, that the $K^{-}pp$ system 
might exhibit a double-pole structure similar to $\Lambda(1405)$. Based 
on the their calculations, such double poles of the $K^{-}pp$ system are 
related to the experimental results. The observed signal close to the 
$\pi\Sigma{N}$ threshold in DISTO and J-PARC E27 experiments, which 
indicate a deeply bound $K^{-}pp$ state are regarded as the second pole 
of the $K^{-}pp$ system, while the observed signal close to the $\bar{K}NN$ 
threshold in J-PARC E15 experiment is considered as the first pole.
The position of a quasi-bound state in the three-body problem is usually 
defined by solving the homogeneous integral equations (\ref{eq10}). To 
find the resonance energy of the three-body system using these equations, 
one should search for a complex energy at which the first eigenvalue of 
the kernel matrix becomes equal to one. The essence of the calculation 
scheme is the integration in the complex plane~\cite{ikeda1}. In the present 
work, we used another way to find the $K^{-}pp$ pole position(s) without 
integration in the complex momentum plane. The signal of the quasi-bound 
state would be observed in the Faddeev amplitudes. 

\begin{figure}[H]
\vspace{-1.5cm}
\hspace{0cm}
\centering
\includegraphics[scale=0.75]{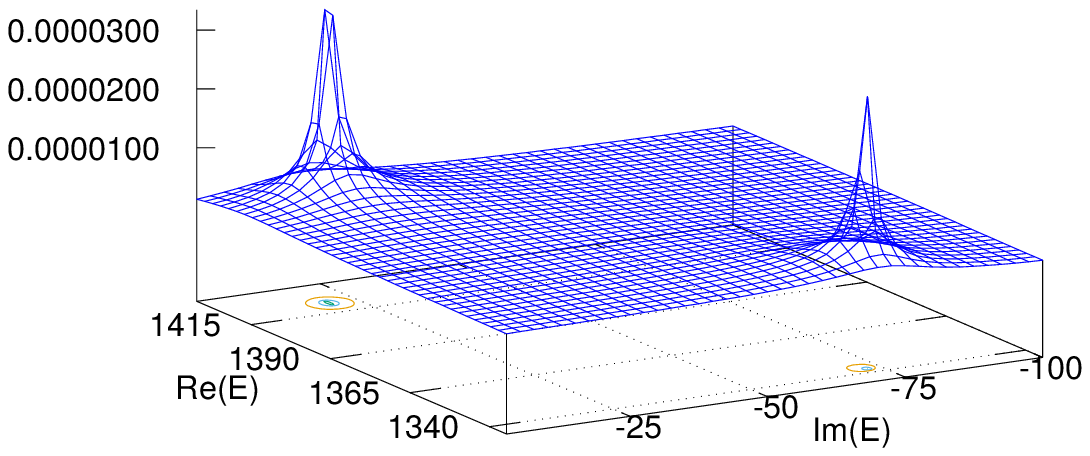} \vspace{-2.cm}\\
\includegraphics[scale=0.75]{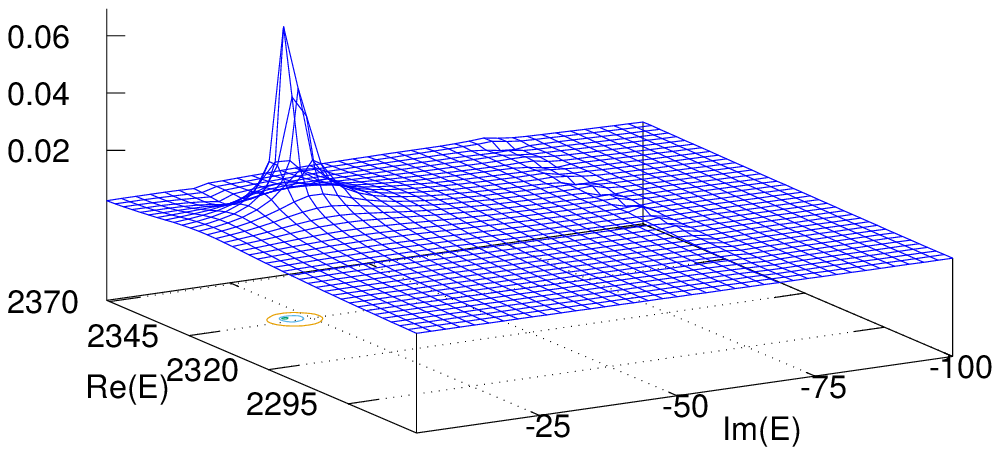} \vspace{-1.cm}
\caption{(Color online) Global view of the calculated scattering amplitudes 
for the two-body $\Bar{K}N$ and three-body $[\bar{K}NN]_{s=0}$ systems, where 
$\bar{K}NN$ is the subsystem of $\bar{K}NNN$ in $\alpha=2$ channel. The upper 
diagram shows the results for $T_{I=0}^{\bar{K}N-\bar{K}N}$ and the lower shows 
the results for $|\mathcal{K}^{2,0}_{NN,00}(q,q',\epsilon)|$ using the same potential, 
where $|\mathcal{K}^{2,0}_{NN,00}(q,q',\epsilon)|$ describes the elastic process 
$[(\bar{K}N)_{I=0}+N]_{s=0} \rightarrow [(\bar{K}N)_{I=0}+N]_{s=0}$. In few-body 
calculations, we used energy-dependent chiral potential for $\bar{K}N-\pi\Sigma$ 
interaction which reproduces the two-pole structure of $\Lambda$(1405) resonance.}
\label{fig2}
\end{figure}

\begin{figure}[H]
\vspace{-1.cm}
\hspace{0cm}
\centering
\includegraphics[scale=0.55]{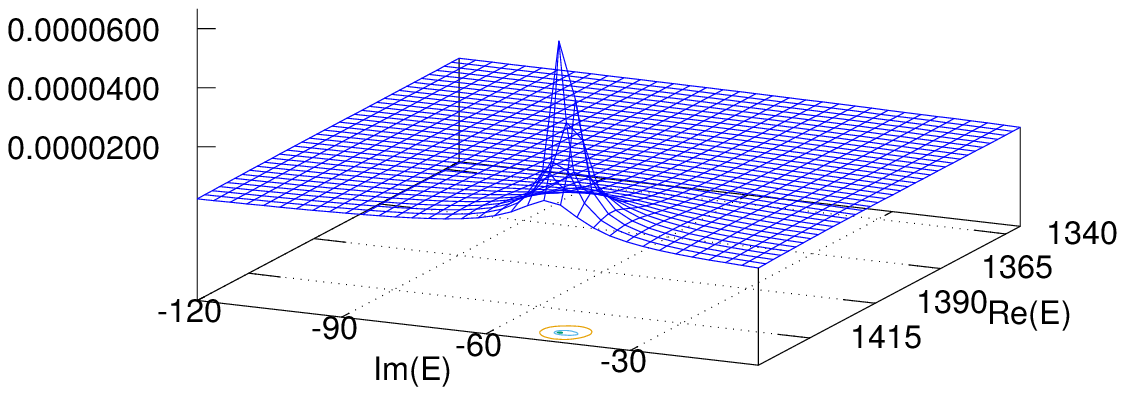} 
\includegraphics[scale=0.55]{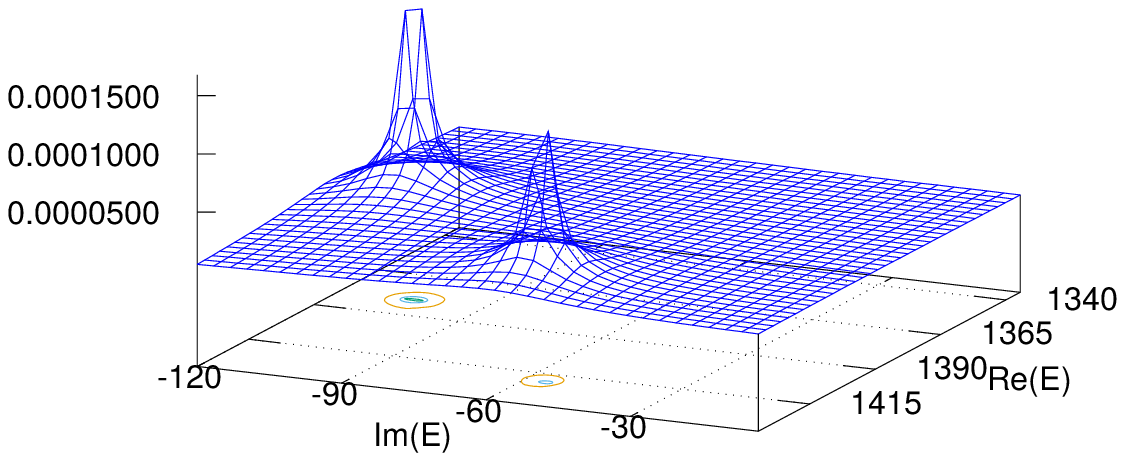} \\
\vspace{-1.cm}
\includegraphics[scale=0.55]{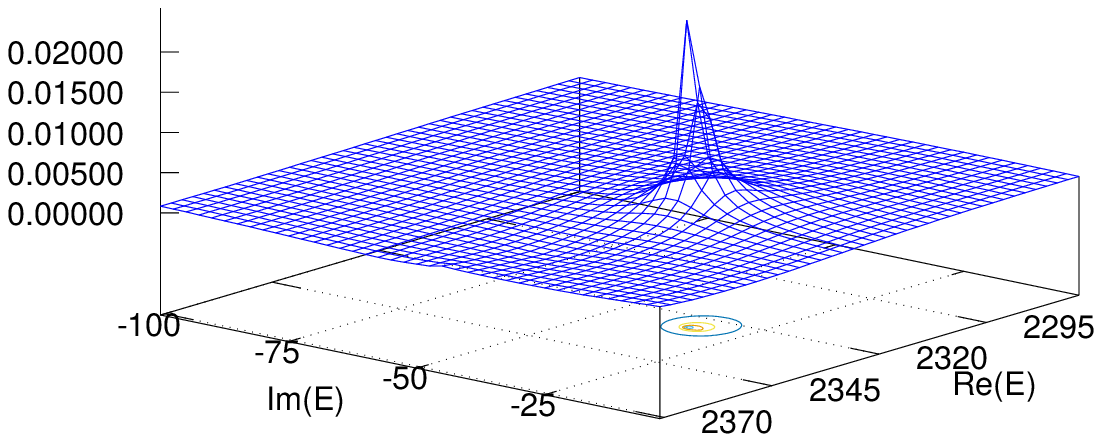} 
\includegraphics[scale=0.55]{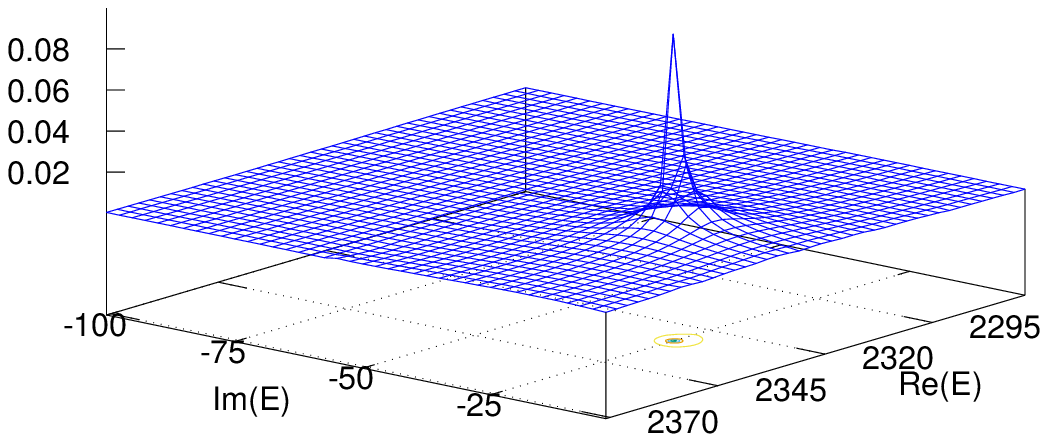} \\
\vspace{-0.cm}
\hspace{0cm}
\caption{(Color online) Same as Fig.\ref{fig2}, but in the present calculations, 
we used the one- and two-pole version of the SIDD potential 
$V^{SIDD}_{\bar{K}N-\pi\Sigma}$. For diagrams on the left side, we used the 
one-pole version and for diagrams on the right side, the two-pole version of 
the SIDD potential was used. The two-body and three-body results are represented 
in the upper and lower row, respectively. }
\label{fig3}
\end{figure}

We studied how the signature of the $K^{-}pp$ system shows up in the 
three-body scattering amplitudes by using coupled-channel Faddeev AGS 
equations. To achieve this goal, we must solve the inhomogeneous integral 
equations for the amplitudes defined in Eq.~(\ref{eq8}). Since the input 
energy of AGS equations is complex the standard moving singularities that 
are caused by the opened channel $\pi\Sigma{N}$, will not appear. With 
this method, we computed the scattering amplitudes at complex energies. 
The calculated resonance energies that have presented in Table \ref{ta1}, 
give pole positions of the $[\bar{K}N]_{I=0}$ and $K^{-}pp$ system and the 
results for Faddeev amplitudes are depicted in Figs.~\ref{fig2} and 
\ref{fig3}. Using Eq.~\ref{eq8}, the amplitude 
$|\mathcal{K}^{2,0}_{NN,00}(q,q',\epsilon)|$ for $(\bar{K}NN)_{s=0}$ 
system is calculated. The operator $|\mathcal{K}^{2,0}_{NN,00}(q,q',\epsilon)|$ is 
the usual Faddeev amplitude, describing the elastic process 
$[(\bar{K}N)_{I=0}+N]_{s=0} \rightarrow [(\bar{K}N)_{I=0}+N]_{s=0}$. In the 
present calculations, the momentums $q$ and $q'$ were taken to be 150MeV/c. 
The real part of the three-body energy, $\epsilon$, changes from 2270 MeV 
to 2370 MeV and the imaginary part changes from -100 to 0 MeV. In Fig.~\ref{fig2}, 
we used the energy-dependent chiral potential to calculate the scattering 
amplitudes and in Fig.~\ref{fig3}, we used one- and two-pole version of the 
SIDD potential. As one can see, both of the poles related to the structure of 
$\Lambda(1405)$ resonance can be seen in two-body scattering amplitudes. One 
close to the $\bar{K}N$ threshold with small width and the other close to the 
$\pi\Sigma$ threshold with large width, while the second pole in the $K^{-}pp$ 
system cannot be seen for all models of $\bar{K}N$ interaction.

\begin{figure}[H]
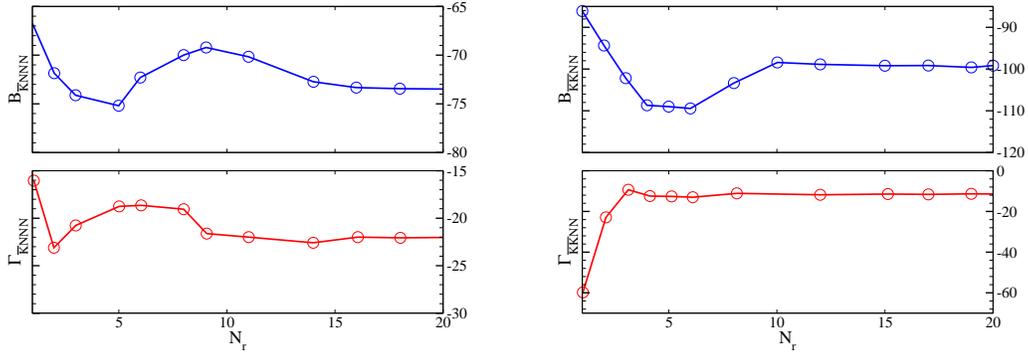

\vspace{0cm}
\hspace{0.cm}
\centering
\includegraphics[scale=0.3]{number1.eps} 
\hspace{1.cm}
\includegraphics[scale=0.3]{number2.eps}
\vspace{-0.0cm}
\caption{(Color online) Diagrammatic representation of dependence of 
the $K^{-}ppn$ and $K^{-}K^{-}pp$ binding energy and width to the 
number of terms ($N_{r}$) in equation (\ref{eq9}). The results say that 
the choice of $N_{r}=15$ will give a reasonable approximation for the 
subamplitudes.}
\label{fig4}
\end{figure}

Starting from Faddeev AGS equations~\ref{eq3} and using different 
versions of the $\bar{K}N-\pi\Sigma$ potentials, the binding energy 
and width of the $K^{-}ppn$ and $K^{-}K^{-}pp$ quasi-bound state were 
evaluated. The dependence of the pole energy on different models of 
$\bar{K}N-\pi\Sigma$ interaction was studied. The separable expansion 
of the Faddeev amplitudes plays an important role, which enable us to 
reduce the four-body Faddeev amplitudes to a single variable integral 
equation. An important parameter in the separable expansion of the Faddeev 
amplitudes is the number of terms ($N_{r}$) in Eq.~\ref{eq9}. In 
Fig.~\ref{fig4}, the sensitivity of the binding energy and width of 
four-body systems to the number of terms $N_{r}$ is investigated. The 
rate of convergence of $K^{-}ppn$ and $K^{-}K^{-}pp$ binding energies 
is investigated. One can see that the choice $N_{r}=15$ provides rather 
satisfactory accuracy.
\begin{table}[H]
\caption{Pole position(s) (in MeV) are extracted from the scattering 
amplitudes. The pole position is related to a quasi-bound states in 
the $\bar{K}N$. The calculations are performed with the SIDDHARTA and 
chiral energy-dependent potential.}
\centering
\begin{tabular}{ccc}
\hline\hline\noalign{\smallskip}
\,\,\, & first pole  \,\,\, &  second pole   \\
\noalign{\smallskip}\hline\noalign{\smallskip}
 $V^{SIDD-1}_{\bar{K}N-\pi\Sigma}$     \,\,\, &  \,\,\, 
$1428.6-i46.5$ \,\,\, &       \\
\noalign{\smallskip}
 $V^{SIDD-2}_{\bar{K}N-\pi\Sigma}$     \,\,\, &  \,\,\, 
$1419.6-i56.0$ \,\,\, &  \,\,\, $1380.1-i104.5$     \\
\noalign{\smallskip}
 $V^{chiral}_{\bar{K}N-\pi\Sigma}$ \,\,\, &  \,\,\, 
$1420.6-i20.3$ \,\,\, &  \,\,\, $1343.0-i72.5$       \\
\noalign{\smallskip}\hline\hline\noalign{\smallskip}
\end{tabular}
\label{ta1} 
\end{table}

In Table~\ref{ta3}, the pole position of the quasi-bound states in 
the $\bar{K}NNN$ and $\bar{K}\bar{K}NN$ systems are presented for 
one- and two-pole version of the SIDDHARTA model of the $\bar{K}N-\pi\Sigma$ 
interaction. The pole energies of the $K^{-}ppn$ and $K^{-}K^{-}pp$ 
systems are calculated with respect to the threshold of the corresponding 
four-body system and by keeping 15 terms in the energy-dependent pole 
expansion of the amplitudes (\ref{eq9}).

\begin{table}[H]
\caption{Pole position(s) (in MeV) of the scattering amplitudes, 
which is related to a quasi-bound state in the $\bar{K}NN$ system. 
The Faddeev AGS calculations for $\bar{K}NN$ system performed with 
the SIDDHARTA and chiral energy-dependent potential. The potentials 
produce the one- and two-pole structure of the $\Lambda$(1405) 
resonance. To extract the results in the second column of the table 
(Direct pole search), we solved Eq.~\ref{eq11} and for driving the 
results in the third column (Faddeev amplitudes), we solved the 
inhomogeneous Faddeev equations~\ref{eq8}.}
\centering
\begin{tabular}{ccc}
\hline\hline\noalign{\smallskip}
\,\,\,\,  & \,\,\,\, Direct pole search \,\,\,\,  
&  \,\,\,\, Faddeev amplitudes   \\
\noalign{\smallskip}\hline\noalign{\smallskip}
$V^{SIDD-1}_{\bar{K}N-\pi\Sigma}$       \,\,\,\,  &  \,\,\,\, 
$2326.0-i34.2$ \,\,\,\,  &  $2326.1-i34.2$   \\
\noalign{\smallskip}
$V^{SIDD-2}_{\bar{K}N-\pi\Sigma}$       \,\,\,\,  &  \,\,\,\, 
$2325.0-i24.1$ \,\,\,\,  &  $2324.5-i24.5$   \\
\noalign{\smallskip}
$V^{chiral}_{\bar{K}N-\pi\Sigma}$   \,\,\,\,  &  \,\,\,\, 
$2346.5-i22.0$ \,\,\,\,  &  $2346.3-i22.0$      \\
\noalign{\smallskip}\hline\hline\noalign{\smallskip}
\end{tabular}
\label{ta2} 
\end{table}

The $\bar{K}N$ channel is strongly coupled to the $\pi\Sigma$ 
channel. Therefore, in actual calculation the $\bar{K}NNN$ and 
$\bar{K}\bar{K}NN$ four-body equations should be generalized to include 
the coupled channels $\bar{K}NNN-\pi\Sigma{NN}$ and 
$\bar{K}\bar{K}NN-\pi\bar{K}\Sigma{N}-\pi\pi\Sigma\Sigma$, respectively. 
Plus the $\bar{K}N-\pi\Sigma$ and nucleon-nucleon interactions, there 
are other interactions in the lower-lying four-body channels, namely 
$\pi\pi$, $\pi\bar{K}$, $\pi{N}$, $\Sigma\bar{K}$, $\Sigma\Sigma$ and 
$\Sigma{N}$ interactions. There is scarce information about some of 
these interactions and also when we include these remaining interactions 
the number of channels will increase rapidly and the treatment of the 
four-body turns out to be very complicated. These computational costs 
can be reduced by using an effective single-channel $\bar{K}N(-\pi\Sigma)$ 
potential. Therefore, in our calculations the lower-lying four-body 
channels are included effectively and consequently the remaining 
interactions in the lower four-body channels are neglected for the systems 
under consideration. Using Eqs.~\ref{eq13} and \ref{eq14}, we can define 
the optical t-matrices in the form
\begin{equation}
T^{I}_{\alpha\alpha}=\frac{1}{1-\lambda^{I,opt}_{\alpha\alpha}G_{\alpha}}
\lambda^{I,opt}_{\alpha\alpha},
\label{eq15}
\end{equation}
where the operator $G_{\alpha}$ is the Green's function in $\alpha$ 
channel and the operator $\lambda^{I,opt}_{\alpha\alpha}$ can be defined by
\begin{equation}
\lambda^{I,opt}_{\alpha\alpha}=\lambda^{I}_{\alpha\alpha}+\lambda^{I}_{\alpha\beta}
\frac{G_{\beta}}{1-\lambda^{I}_{\beta\beta}G_{\beta}}\lambda^{I}_{\beta\alpha},
\label{eq16}
\end{equation}

A definitive study of the $K^{-}ppn$ and $K^{-}K^{-}pp$ bound 
states could be performed using standard energy-dependent $\bar{K}N$ 
input potential, too~\cite{ikeda4}. The energy-dependent potentials 
provide a weaker $\bar{K}N$ attraction for lower energies than the 
energy-independent potentials. Therefore, one expects that, the 
quasi-bound states resulting from the energy-dependent potential happen 
to be shallower. The comparison of the obtained results for the chiral 
$\bar{K}N-\pi\Sigma$ interaction with the calculated binding energies 
for phenomenological $\bar{K}N$ interaction shows that energy-independent 
potentials produce much deeper bound state for four-body kaonic systems 
under consideration in the present work.

It was shown in Ref.~\cite{shev4}, that exact optical potential approximation is more 
accurate for the one-pole version of the $\bar{K}N$ interaction than for the two-pole 
model of interaction. In other words, the optical approximation provides exactly the 
same elastic $\bar{K}N$ amplitude as the coupled-channel model of interaction for one-pole 
potential while for two-pole model, we cannot see this behavior. In present calculations, 
we can expect that the binding energies resulting from the optical approximation of the 
$V^{2,SIDD}_{\bar{K}N-\pi\Sigma}$ potential be different from the full coupled channel 
calculations of the $\bar{K}NNN-\pi\Sigma{NN}$ and $\bar{K}\bar{K}NN-\pi\pi\Sigma\Sigma$ 
systems. Comparing the results in Tables~\ref{ta2} and \ref{ta3}, we can see that the 
binding energy of the $K^{-}pp$ system resulting from both one- and two-pole models of 
the SIDD potential are close to each other while in four-body calculations, the one-pole 
potential reproduce a deeper bound state for $K^{-}ppn$ and $K^{-}K^{-}pp$ systems. This 
difference may come from the fact that the optical approximation is not very appropriate 
for two-pole potentials. The fully coupled-channel calculations of the systems under 
consideration in the future may help us to make a better judgment.

The calculated binding energy and width values of the $K^{-}ppn$ and 
$K^{-}K^{-}pp$ quasi-bound state are compared in Table~\ref{ta3} with 
other theoretical results. The Faddeev calculations of the $K^{-}ppn$ 
and $K^{-}K^{-}pp$ systems were also carried out in Refs.~\cite{kh1,kh2} 
using the same $V_{\bar{K}N}$ potentials. The HSE method was used there 
to find the separable expression of the Faddeev amplitudes in (2+2) and 
(3+1) subsystems. Comparing the present results for $K^{-}ppn$ and 
$K^{-}K^{-}pp$ systems with those in~\cite{kh1} and~\cite{kh2} shows that 
the obtained binding energies within the EDPE method are deeper than those 
resulting from the HSE method.

Faddeev-Yakubowsky equations were solved in~\cite{maeda1} with phenomenological 
energy independent $\bar{K}N$ potentials. Therefore, in principle, their 
calculation with the energy independent version of the $\bar{K}N$ potential 
should give a result, which are close to ours with a phenomenological model 
of interaction. The variational calculations using AY potential were also carried 
out in~\cite{roman}. It is seen, however, that only their binding 
energies are comparable to ours, while the widths obtained in~\cite{roman} 
are much bigger than ours. Variational calculations using chiral energy 
independent $\bar{K}N$ potential were also done in Refs.~\cite{gal1,roman}. 
The obtained binding energies are shallower than those calculated in 
the present work. It is caused by the relative weakness of the chiral $\bar{K}N$ 
interaction as compared to phenomenological $\bar{K}N$.
\begin{table}[H]
\caption{The dependence of the pole position(s) (in MeV), of the 
$K^{-}ppn$ and $K^{-}K^{-}pp$ systems on the different models of 
the $\bar{K}N-\pi\Sigma$ interactions is investigated. 
$V^{SIDD-1}_{\bar{K}N-\pi\Sigma}$ and $V^{SIDD-2}_{\bar{K}N-\pi\Sigma}$ 
standing for a one-pole and a two-pole structure of the $\Lambda$(1405) 
resonance, which are produced phenomenologically and 
$V^{chiral}_{\bar{K}N-\pi\Sigma}$ is used for energy-dependent chiral 
potential. The calculated binding energies and widths are also compared 
with other theoretical results. It was referred to the potentials in 
Refs.~\cite{akaishi} and ~\cite{hyodo} as the Akaishi-Yamazaki (AY) and 
HW potentials, respectively. The first one is based on chiral dynamics 
and the second one is constructed phenomenologicaly.}
\centering
\begin{tabular}{ccccc}
\hline\hline\noalign{\smallskip}\noalign{\smallskip}
 \,\,\, & $B_{K^{-}ppn}$ \,\,\, & \,\,\, $\Gamma_{K^{-}ppn}$ \,\,\, 
& \,\,\, $B_{K^{-}K^{-}pp}$ \,\,\, 
& \,\,\, $\Gamma_{K^{-}K^{-}pp}$ \,\,\, \\
\noalign{\smallskip}\noalign{\smallskip}
\hline
\noalign{\smallskip}\noalign{\smallskip}
Present AGS (EDPE): \,\,\, &\,\,\,  \,\,\, &\,\,\,  \,\,\, &\,\,\,  
\,\,\, &\,\,\,   \\
\noalign{\smallskip}\noalign{\smallskip}
with $V^{1,SIDD}_{\bar{K}N-\pi\Sigma}$ \,\,\, &\,\,\,  $73.5$ \,\,\, &\,\,\,  
$22.0$ \,\,\, &\,\,\,  $99.2$ \,\,\, &\,\,\,  $11.4$ \\
\noalign{\smallskip}\noalign{\smallskip}
with $V^{2,SIDD}_{\bar{K}N-\pi\Sigma}$ \,\,\, &\,\,\,  $58.5$ \,\,\, &\,\,\,  
$27.0$ \,\,\, &\,\,\,  $89.0$ \,\,\, &\,\,\,  $11.4$ \\
\noalign{\smallskip}\noalign{\smallskip}
with $V^{chiral}_{\bar{K}N-\pi\Sigma-\pi\Lambda}$ \,\,\, &\,\,\,  $41.4$ \,\,\, 
&\,\,\,  $31.5$ \,\,\, &\,\,\,  $60.9$ \,\,\, &\,\,\,  $65.0$ \\
\noalign{\smallskip}\hline\noalign{\smallskip}
Previous AGS (HSE): \,\,\, &\,\,\,  \,\,\, &\,\,\,  \,\,\, &\,\,\,  \,\,\, &\,\,\,   \\
\noalign{\smallskip}\noalign{\smallskip}
with $V^{1,SIDD}_{\bar{K}N-\pi\Sigma}$ \,\,\, &\,\,\,  $68.8$~\cite{kh1} \,\,\, 
&\,\,\,  $22.0$~\cite{kh1} \,\,\, &\,\,\,  $93.7$~\cite{kh2} \,\,\, &\,\,\,  $30.6$~\cite{kh2} \\
\noalign{\smallskip}\noalign{\smallskip}
with $V^{2,SIDD}_{\bar{K}N-\pi\Sigma}$ \,\,\, &\,\,\,  $55.9$~\cite{kh1} \,\,\, 
&\,\,\,  $17.6$~\cite{kh1} \,\,\, &\,\,\,  $84.2$~\cite{kh2} \,\,\, &\,\,\,  $7.8$~\cite{kh2} \\
\noalign{\smallskip}\noalign{\smallskip}
with $V^{1,KEK}_{\bar{K}N-\pi\Sigma}$~\cite{kh2} \,\,\, &\,\,\,  $-$ \,\,\, 
&\,\,\,  $-$ \,\,\, &\,\,\,  $84.6$ \,\,\, &\,\,\,  $24.2$ \\
\noalign{\smallskip}\noalign{\smallskip}
with $V^{2,KEK}_{\bar{K}N-\pi\Sigma}$~\cite{kh2} \,\,\, &\,\,\,  $-$ \,\,\, 
&\,\,\,  $-$ \,\,\, &\,\,\,  $81.8$ \,\,\, &\,\,\,  $4.6$ \\
\noalign{\smallskip}\hline\noalign{\smallskip}
Faddeev-Yakubowsky: & & & &  \\
\noalign{\smallskip}\noalign{\smallskip}
MAY~\cite{maeda1} & $74$ & $-$ & $104$ & $-$ \\
\noalign{\smallskip}\hline\noalign{\smallskip}
Variational: & & & & \\
BGL~\cite{gal1} & $29.3$ & $32.9$ & $32.1$ & $80.5$ \\
\noalign{\smallskip}\noalign{\smallskip}
KTT with AY~\cite{akaishi} & $92-98$ & $\sim 83$ & $\sim 92$ & $\sim 73$ \\
\noalign{\smallskip}\noalign{\smallskip}
KTT with HW~\cite{hyodo} & $\sim 29$ & $\sim 30$ & $\sim 32$ & $\sim 79$ \\
\noalign{\smallskip}\noalign{\smallskip}
\hline\hline
\end{tabular}
\label{ta3} 
\end{table}
\section{CONCLUSION}
\label{conc}
In summary, the Faddeev-type calculations of $\bar{K}NNN$ and $\bar{K}\bar{K}NN$ 
systems were performed. We have calculated the binding energy and width of these 
kaonic systems. To investigate the dependence of the resulting binding energies 
and widths on models of $\bar{K}N-\pi\Sigma$ interaction, different versions of 
$\bar{K}N-\pi\Sigma$ potentials, which produce the one- or two-pole structure of 
$\Lambda$(1405) resonance, were used. In the present calculations, we approximated 
the full coupled-channel one- and two-pole models of interaction by constructing 
the exact optical $\bar{K}N-\pi\Sigma$ potential. Therefore, one-channel Faddeev 
AGS equations are solved for the $\bar{K}NNN$ and $\bar{K}\bar{K}NN$ systems and 
the decaying to the $\pi\Sigma{NN}$ and $\pi\pi\Sigma\Sigma$ channels is taken into 
account through the imaginary part of the optical $\bar{K}N(-\pi\Sigma)$ potential. 
For $K^{-}ppn$ system, we obtained binding energy $\sim 41$ MeV using the chiral 
and $58-73$ MeV for the SIDD $\bar{K}N$ potentials. The width is about $\sim 30$ 
MeV for chiral potential, while the SIDD potentials give $\sim 22-27$ MeV. The 
calculations yielded binding energy $B_{chiral}\sim$ 61 and $B_{pheno.}\sim$ 90-100 
MeV for $K^{-}K^{-}pp$ system. The obtained widths for $K^{-}K^{-}pp$ are 
$\Gamma_{chiral}\sim$ 65 and $\Gamma_{pheno.}=11$ MeV. It is expected that the  of 
The omission of the lower-lying four-body channels may has an important effect on 
the width of the state specially in the case of $\bar{K}\bar{K}NN$ system. The 
$\pi\pi\Sigma\Sigma$ threshold is much lower in comparison to the bound state as 
the one for $\pi\Sigma{NN}$ in the case of $\bar{K}NNN$ system. Thus, there is more 
phase space available for the decay and that again could lead to a larger width. 
Therefore, the full coupled-channel calculations of the systems under consideration 
in the future may help us to make a better judgment about the effects of the lower-lying 
channels. plus the four-body systems, the $\bar{K}NN$ system was also studied. The 
Faddeev amplitudes for this system was calculated and it was shown that double-pole 
structure cannot be seen in the Faddeev amplitudes.
\section{Acknowledgements}
This work has been financially supported by the research deputy of Shahrekord University. 
The grant number was 141/1530.


\begin{thebibliography}{99}
\bibitem{dal1} R. H. Dalitz, S. F. Tuan, Phys. Rev. Lett. {\bf 2}, 425 (1959). 
\bibitem{dal2} R. H. Dalitz, S. F. Tuan, Annals Phys. {\bf 10}, 307 (1960). 
\bibitem{akaishi} Y. Akaishi and T. Yamazaki, Phys. Rev. C {\bf 65}, 044005 (2002).
\bibitem{esm2}  J. Esmaili, Y. Akaishi and T. Yamazaki, Phys. Lett. B {\bf 686}, 23-28 (2010).
\bibitem{esm1} J. Esmaili,Y. Akaishi and T. Yamazaki, Phys. Rev. C {\bf 83}, 055207 (2011).
\bibitem{yamazaki1} T. Yamazaki and Y. Akaishi, Phys. Lett. B {\bf 535}, 70 (2002).
\bibitem{dot1} A. Dote, H. Horiuchi, Y. Akaishi and T. Yamazaki, Phys. Lett. B {\bf 590}, 51 (2004).
\bibitem{dot2} A. Dote, H. Horiuchi, Y. Akaishi and T. Yamazaki, Phys. Rev. C {\bf 70}, 044313 (2004).
\bibitem{shev1} N. V. Shevchenko, A. Gal and J. Mares, Phys. Rev. Lett. {\bf 98}, 082301 (2007).
\bibitem{shev2} N. V. Shevchenko, A. Gal, J. Mares and J. Revai, Phys. Rev. C {\bf 76}, 044004 (2007).
\bibitem{ikeda1} Y. Ikeda and T. Sato, Phys. Rev. C {\bf 76}, 035203 (2007).
\bibitem{ikeda2} Y. Ikeda and T. Sato, Phys. Rev. C {\bf 79}, 035201 (2009).
\bibitem{dote1} A. Dote, T. Hyodo and W. Weise, Nucl. Phys. A {\bf 804}, 197 (2008).
\bibitem{dote2} A. Dote, T. Hyodo and W. Weise, Phys. Rev. C {\bf 79}, 014003 (2009).
\bibitem{ikeda3} Y. Ikeda, H. Kamano and T. Sato, Prog. Theor. Phys. {\bf 124}, 533 (2010).
\bibitem{kh1} S. Marri and S. Z. Kalantari, Eur. Phys. J. A {\bf 52}, 282 (2016), arXiv:1611.09025 [nucl-th].
\bibitem{kh2} S. Marri, S. Z. Kalantari and J. Esmaili, Eur. Phys. J. A {\bf 52}, 361 (2016), arXiv:1612.00685 [nucl-th].
\bibitem{kh3} S. Marri, S. Z. Kalantari and J. Esmaili, arXiv:1903.07340 [nucl-th].
\bibitem{maeda} S. Maeda, Y. Akaishi and T. Yamazaki, Proc. Jpn. Acad., Ser. B {\bf 89}, 418 (2013).
\bibitem{gal1} N. Barnea, A. Gal and E. Z. Liverts, Phys. Lett. B {\bf 712}, 132 (2012).
\bibitem{agnel} M. Agnello, {\it et al.}, Phys. Rev. Lett. {\bf 94}, 212303 (2005).
\bibitem{bend} G. Bendiscioli, {\it et al.}, Nucl. Phys. A {\bf 789}, 222 (2007).
\bibitem{yamaz} T. Yamazaki, {\it et al.}, Phys. Rev. Lett. {\bf 104}, 132502 (2010).
\bibitem{aji} S. Ajimura, {\it et al.}, Nucl. Phys. A {\bf 914}, 315 (2013).
\bibitem{ichi} Y. Ichikawa, {\it et al.}, Few Body Syst. {\bf 54}, 1191 (2013).
\bibitem{fabbi} L. Fabbietti, {\it et al.}, Nucl. Phys. A {\bf 914}, 60 (2013).
\bibitem{toki} A. O. Tokiyasu, {\it et al.}, Phys. Lett. B {\bf 728}, 616 (2014).
\bibitem{naro} I. M. Narodetsky, Nucl. Phys. A {\bf 221}, 191 (1974).
\bibitem{sofia} S. A. Sofianos, N. J. McGurk, and H. Fiedeldey, Nucl. Phys. A {\bf 318}, 295 (1979).
\bibitem{dot3} A. Dote, T. Inoue, T. Myo, Phys. Rev. C {\bf 95}, 062201 (2017).
\bibitem{ikeda4} Y. Ikeda, H. Kamano, and T. Sato, Prog. Theor. Phys. {\bf 124}, 533 (2010).
\bibitem{shev3} N. V. Shevchenko, Nucl. Phys. A {\bf 890-891}, 50 (2012).
\bibitem{grass} P. Grassberger, W. Sandhas, Nucl. Phys. B {\bf 2}, 181 (1967).
\bibitem{fonce} A. C. Fonseca, L. Streit, Lect. Notes Phys. {\bf 273}, 161 (1986).
\bibitem{fix} A. Fix and H. Arenhovel, Phys. Rev. C {\bf 68}, 044002 (2003).
\bibitem{naka} S. Nakaichi, T. K. Lim, Y. Akaishi and H. Tanaka, Phys. Rev. A {\bf 26}, 1 (1982).
\bibitem{shev4} N. V. Shevchenko, Phys. Rev. C {\bf 85}, 034001 (2012).
\bibitem{sidd} (SIDDHARTA Collaboration) M. Bazzi {\it et al.}, Phys. Lett. B {\bf 704}, 113 (2011).
\bibitem{pest} H. Zankel, W. Plessas and J. Haidenbauer, Phys. Rev. C {\bf 28}, 538 (1983).
\bibitem{bean} S. R. Beane {\it et al.}, (NPLQCD Collaboration), Phys. Rev. D {\bf 77}, 094507 (2008).
\bibitem{hyodo} T. Hyodo and W. Weise, Phys. Rev. C {\bf 77}, 035204 (2008).
\bibitem{maeda1} S. Maeda, Y. Akaishi and T. Yamazaki, arXiv:1610.02150v1 [nucl-ex].
\bibitem{roman} Roman Ya. Kezerashvili, Shalva M. Tsiklauri and Nurgali Zh. Takibayev, arXiv:1510.00478v3 [nucl-th].
\end{thebibliography}
\end{document}